# Type-II Superconductivity at 9K in Pb-Bi Alloy

N. K. Karn[1,2], Kapil Kumar[1,2], Naveen Kumar[1,2], Yogesh Kumar[1,2], M. M. Sharma[3], Jin Hu[3] and V.P.S. Awana[1,2,*]

[1]Academy of Scientific & Innovative Research (AcSIR), Ghaziabad, 201002, India
[2]CSIR- National Physical Laboratory, New Delhi, 110012, India
[3]Department of Physics, University of Arkansas, Fayetteville, AR 72701, USA

**Abstract**

In the present work, we report the synthesis of Pb-Bi alloy with enhanced $T_c$ of up to 9K, which is higher than that of Pb. The alloy is synthesized via a solid-state reaction route in the vacuum-encapsulated quartz tube at $700^0C$ in an automated furnace. The synthesized sample is characterized by X-ray Diffraction(XRD) and Energy dispersive X-ray analysis(EDAX) for its phase purity and elemental composition. Rietveld refinement of XRD reveals that the end product is a majority hexagonal $Pb_7Bi_3$, with minor rhombohedral Bi. The electronic transport measurement shows metallic behavior with the Debye temperature of 108K and a superconductivity transition temperature ($T_c$) below 9K, which is the maximum to date for any reported Pb-Bi alloy, Pb or Bi at ambient pressure. Partial substitution of Bi at the Pb site may modify the free density of electronic states within the BCS model to attain the optimum $T_c$, which is higher by around 2K from the reported $T_c$ of Pb. The superconductor phase diagram derived from magneto-transport measurements reveals that the synthesized alloy is a conventional superconductor with an upper critical field ($H_{c2}$) of 3.9Tesla, which lies well within the Pauli paramagnetic limit. The magnetization measurements carried out following ZFC(Zero Field Cool) protocols infer that the synthesized alloy is a bulk superconductor below 9K. The isothermal M-H(Magnetization vs. Field) measurements performed below $T_c$ establish it as a type-II superconductor. The specific heat capacity measurements show that the Pb-Bi alloy is a strongly coupled bulk superconductor below around 9K with possibly two superconducting gaps.

*Corresponding Author
Dr. V. P. S. Awana, Chief Scientist
CSIR-National Physical Laboratory, India
E-mail: aawana@nplindia.org
Ph. +91-11-45609357, Fax-+91-11-45609310

**Introduction**

Superconductivity has been one of the most intriguing phenomena since its discovery in 1911 [1]. Among the various superconducting materials, superconductivity in alloys like HgAu, HgCd and PbSn [2] was discovered soon after the superconductivity was found in Hg [1]. The superconducting properties in alloys are tunable and can be engineered by varying alloy compositions [3-6]. Recently, alloys of TaHf and TaZr have been shown to have applications in superconducting devices [7].

Among different alloy systems, lead-bismuth (PbBi) alloy has attained significant interest due to its interesting electronic, topological [8, 9], and superconducting properties [10-12]. The combination of lead and bismuth in these alloys can offer promising avenues for exploring fundamental physics and practical applications, like tuning superconductivity by atomically controlled thin film growth [13]. In this context, our research focuses on synthesizing Pb-Bi alloys via the solid-state reaction method and characterizing their properties [14], particularly investigating their superconducting behavior.

Recent studies have revealed fascinating aspects of Pb-Bi alloys. The transformation in crystal structure and superconducting behavior has been investigated by varying the lead concentration in Pb-Bi alloy [15]. The alloy is found to be an s-wave superconductor with its critical temperature depending upon the weight percentage of both elements in the alloy. Experimental efforts have been put in to explore the transport properties of Pb-Bi thin films grown by thermal evaporation, demonstrating the superconducting transition temperature and its dependence on the applied magnetic field [16].

Pb-Bi alloys are significant because of their possible use in quantum technologies and superconducting devices [4, 17]. Considering the superconducting properties of Pb-Bi alloys, particularly their critical temperature, is crucial for advancing materials science and exploring unconventional superconductivity at lower temperatures.

In this article, we synthesized Pb-Bi alloys using a solid-state reaction method, followed by comprehensive characterization using X-ray diffraction (XRD) to determine their crystal structure. Our transport measurements revealed the superconducting nature of Pb-Bi with a critical temperature ($T_c$) of 9 K. The upper critical field is estimated using GL theory formulation which is found to be higher compared to previous reports. From the magnetization measurements, the wide open M-H loop gives clear evidence of type-II superconductivity confirmed by computing critical superconducting parameters within the Ginzberg-Landau formalism. The bulk superconductivity is evidenced by specific heat capacity measurement with strong electron-phonon (el-ph) coupling. These findings underscore the potential of Pb-Bi alloys for low-temperature superconducting applications and motivates further investigation into their underlying physical mechanisms.

**Experimental**

The alloy of Pb-Bi is prepared in a simple single step by following the solid-state reaction route. High-purity (>4 N) powders of Pb and Bi were taken into the stoichiometric ratio 1:1. These powders were mixed and ground thoroughly by using an agate mortar pestle in an argon-filled glove box. This homogenous mixture was then palletized and vacuum encapsulated at a pressure of $5\times10^{-5}$ mbar. This vacuum-encapsulated sample was then placed into an automated PID-controlled Muffle furnace subjected to the heat treatment shown in Fig.1. The encapsulated pallet is heated to 700°C at a rate of 60°C/h. This sample is kept at this elevated temperature for 48h and then cooled normally to room temperature. As obtained crystalline Pb-Bi alloy is shown in the inset of Fig.1. Rigaku mini flex tabletop X-ray diffractometer (XRD) equipped with Cu-K$_\alpha$ radiation of wavelength 1.54Å is used to record the XRD pattern of the synthesized alloy Pb-Bi. Zeiss EVO-50 made field emission scanning electron microscopy (FESEM) equipped with energy dispersive x-ray analysis (EDAX) was used to visualize the surface morphology and elemental composition, respectively. Electrical transport and Heat capacity measurements have been carried out using a Quantum Design (QD) made Physical Property Measurement System (PPMS) with the sample mounted on a PPMS DC and Heat capacity pucks respectively. A standard four-probe method has been used for the magneto transport measurements, in which four linear contacts with silver epoxy have been made on the surface of the sample. The magnetic properties are elucidated through magnetization measurements performed on the MPMS SQUID magnetometer.

**Results and Discussion**

After heat treatment, the obtained crystals are silvery shiny and around 1cm long as shown in the inset of Fig. 1. Fig. 2(a) exhibits the Rietveld refined powder XRD (PXRD) pattern of synthesized Pb-Bi alloy which fits in two phases $Pb_7Bi_3$ and Bi. The $Pb_7Bi_3$ crystallizes in a hexagonal phase with P63/mmc space group, whereas Bi has a rhombohedral structure with space group R-3m. The goodness of fit parameter, i.e., $\chi 2$ is found to be 3.7 which is reasonable. The obtained lattice parameters are a=b= 3.5051(1)Å and c= 5.7984(2)Å, and α=β=90, γ=120º for $Pb_7Bi_3$ and the same for Bi is obtained as a=b= 4.5435(9)Å and c= 11.8694(5)Å and α=β=90, γ=120º. The obtained lattice parameters are found close to the previous reports [15, 18]. The results of powder XRD refinement are summarized in Table 1. All the observed peaks in the PXRD pattern are spaned by the Bragg positions generated by the two phases. Corresponding to the observed peaks, significant planes are marked for both $Pb_7Bi_3$ (Black color) and Bi(Green color). Fig. 2(b) shows the VESTA-drawn unit cell of both phases present in the alloy with the refined lattice parameters shown in table 1.

Fig. 2(c) shows the EDAX spectra of the studied Pb-Bi alloy illustrating the observed characteristic peaks of Pb and Bi. The inset table shows the atomic and weight percentages of constituent elements Pb and Bi with 54.41% and 45.59% weight percentages, respectively. Following the XRD refinement results which show a two-phase $Pb_7Bi_3$ and Bi sample, the EDAX weight percentage analysis results in the sample composition as $(Pb_7Bi_3)_{0.262}Bi_{0.738}$. The minor peaks observed at low energies can be attributed to the addendum elements Carbon and Oxygen which are deposited to the surface due to the exposure of the sample to the atmosphere. The inset image of Fig. 2(c) shows the SEM image of the Pb-Bi alloy at a resolution of 2μm. For single crystals, the SEM image usually shows layers, stairs or terrace-type morphology. But here, the absence of such features and the presence of irregular shapes at the surface confirm the polycrystalline morphology of the sample.

Next, the electronic and magneto-transport properties of the as-synthesized Pb-Bi alloy have been studied by performing transport measurements on PPMS. Fig. 3(a) shows the resistivity vs temperature (ρ-T) plot at zero magnetic field. A superconducting transition is observed (see inset of Fig. 3(a) as the resistivity starts decreasing sharply at $T_c^{onset}$ = 9±0.2K, which is close to the previously reported values [15, 19]. The temperature at which ρ =0 ($T_c^{offset}$) is seen, is at 8.7±0.2K, with a transition width of ~0.3K. Above $T_c$, the resistivity increases with temperature, showing high metallicity of the synthesized crystal. Above $T_c$ (i.e., 9K–300K) the ρ–T data are fitted using the Bloch–Grüneisen (B-G) formula given below

$$\rho(T) = \left[\frac{1}{\rho_s} + \frac{1}{\rho_i(T)}\right]^{-1} \qquad (1)$$

Here, $\rho_s$ denotes the temperature-independent saturation resistivity [20] and $\rho_i(T)$ is the temperature-dependent term which can be given by:

$$\rho(T) = \rho(0) + \rho_{el-ph}(T) \tag{2}$$

Here, $\rho(0)$ denotes residual resistivity arising due to impurity scattering and $\rho_{el-ph}(T)$ denotes the temperature-dependent term, which depends on electron-phonon scattering. Furthermore, $\rho_{el-ph}(T)$ is given by the following formula:

$$\rho_{el-ph} = \alpha_{el-ph} \left(\frac{T}{\theta_D}\right)^n \int_0^{\frac{\theta_D}{T}} \frac{x^n}{(1-e^{-x})*(e^x-1)} dx \tag{3}$$

Here, $\alpha_{el-ph}$ is the electron-phonon coupling parameter, $\theta_D$ represents the Debye temperature, and n is a constant. The $\rho$–T data are well fitted for n = 5, signifying dominant electron-phonon scattering above $T_c$. The deduced values are $\theta_D$ =107.72K and $\rho(0)$ = 0.159μΩ-cm. The residual resistivity ratio (RRR) given by $\rho(300K)/\rho(T_c)$ is ~1.88. The resistivity of the synthesized alloy is of the order of $10^{-7}$Ω-cm which is lower than Copper. This endorses the quality and the high metallicity of the synthesized Pb-Bi alloy.

To study the upper critical field and other superconducting parameters of the synthesized alloy, magnetotransport measurements have been performed. Fig. 3(b) depicts ρ-T plots in the temperature range 2K-15K at different fields. We observe that, for low fields, the superconductor transition is sharp but at higher fields, the transition width increases. Fig. 3(c) illustrates the resistivity vs field ρ-H plot with applied field ±6Tesla for different temperatures 2K to 10K at the interval of 0.5K. Similar to ρ-T plots, at lower temperatures, the gap in $H_{c2}^{onset}$ and $H_{c2}^{offset}$ increases.

Further, the superconducting properties are elucidated by the DC isothermal magnetization M-H and M-T measurements following ZFC and FC protocols. Fig. 4(a) shows the ZFC and FC branches of the M-T curve with the onset $T_c$ of 9.1K at 10Oe, which is within the range of uncertainties and in agreement with the enhanced $T_c$ observed in resistivity measurements in comparison to previous reports [15]. The M-T measurement in ZFC mode shows a sharp diamagnetic transition in the synthesized Pb-Bi alloy. Interstingly, Bi is the most diamagnetic material among all elements [21] and the synthesized alloy has a Bi phase as well. However, the diamagnetic response is feeble compared to a superconductor with the order difference of ~$10^{-4}$.

Thus, the observed diamagnetic transition is only due to the superconducting phase $Pb_7Bi_3$ present in the synthesized alloy. In the low field limit, the magnetic susceptibility is obtained simply by $\chi=M/H$. Using the M-T data, the diamagnetic susceptibility is obtained and shown in the inset of Fig. 4(a). The ZFC susceptibility ($-4\pi\chi$) represents the shielding fraction in a superconducting material [22], which is found to be ~0.1 at 2K. This low shielding fraction can be attributed to the small superconducting $Pb_7Bi_3$ phase present in the Pb-Bi alloy as shown by EDAX and XRD characterization.

Gandhi et. al. [15] have done an extensive study of Pb-Bi alloy focused on magnetic characterization, but they do not report any FC measurement. In Fig. 4(a), the FC measurements show the upturn transition below $T_c$, showing the paramagnetic Meissner effect (PME). It is different from the transition usually observed for a superconductor [23-25], as the diamagnetism below Tc is supposed to be the hallmark of superconductivity. Several explanations of PME have been proposed, which can be broadly classified as (i)Extrinsic PME which originates from sample surface superconductivity [26], vortex-vortex interaction [27] (ii)Intrinsic PME: where PME is a inherent property of the material that includes the formation of $\pi$ junctions [28], s-wave odd frequency superconductivity two-band superconductor [29]. The PME effect may be observed due to the presence of paramagnetic impurities. In the synthesized Pb-Bi alloy, apart from the superconducting phase $Pb_7Bi_3$, Bi is present, but Bi itself is diamagnetic. So, the origin of the observed Paramagnetic Meissner Effect (PME) can be attributed to the presence of the interface of the metal-superconductor explained with conventional superconductivity [30]. Here, the paramagnetic Meissner effect is consistent with the results of the ZFC measurements as it shows clear bifurcation of superconducting transition onset ~9K. To get more insight into the observed PME, further investigations such as time relaxation magnetic measurements and others are warranted.

Fig. 4(b) shows the hysteresis curves in the five quadrants for the Pb-Bi alloy at temperatures 2K and 5K, below $T_c$ characterized by symmetric M-H loops. The wide open butterfly-like symmetric loops in the M-H plots are the signature plots of any type-II superconductor and thereby it confirms Type-II superconductivity in Pb-Bi alloy. The linear response of magnetization in the first quadrant is evident from the inset of Fig. 4(b), which can be attributed to the clear regime of Miessner state. Lower critical field $H_{c1}$ is defined as the field at which the M-H plot starts to deviate from the linearity and enters into the vortex state from the Meissner state. Once $H_{c1}$ crosses, the M-H loop becomes highly irreversible. There are several methods proposed by different groups to determine the $H_{c1}$ [31-33]. Here, we follow the simplest method: $H_{c1}$ is the point from which the M-H curve starts deviating from the linearity. Since, the actual magnetic field around the sample is larger than the applied magnetic field due to the demagnetization effects by a factor of $1/(1 - N)$, where N is the demagnetization factor. We have to correct the $H_{c1}$ values for such an effect [34]. Following the method proposed by Brandt [35], the demagnetization factor is calculated by the formula $N = 1 - 1/\left(1 + q\frac{a}{b}\right)$, where

$$q = \frac{\pi}{4} + 0.64 \tanh\left[0.64\frac{b}{a}\ln\left(1.7 + 1.2\frac{a}{b}\right)\right] \tag{4}$$

is for slab type samples. Here, for the measured sample $a = 2$mm and $b = 5.8$mm resulting $q = 1.35$. The demagnetization factor (N) is found to be 0.317. After demagnetization correction, the $H_{c1}$ is found to be 273Oe, 205Oe and 68Oe at 2K, 5K and 8K respectively. The inset of Fig. 5 depicts the variation of lower critical field $H_{c1}$ with the reduced temperature $t = T/T_c$, for the data extracted from the M-H plot. Here T is the temperature at which the experiment is carried out. To obtain the lowest critical field $H_{c1}(0)$, the data is fitted with the Gingberg-Landau model [36]:

$$H_{c1}(T) = H_{c1}(0)[1 - t^2] \tag{5}$$

and extrapolated to 0K. The fitted and extrapolated curve is shown in the inset of Fig. 5 in Blue color which results in the lower critical field at 0K, $H_{c1}(0) = 291.6$Oe.

Using the data from Fig. 4(a) and (b), the superconducting phase diagram is produced in the phase space of ($T_c$, $H_{c2}$) as shown in Fig. 5. From the phase diagram, it is evident that the $\Delta H_{c2}$ =$H_{c2}^{onset} - H_{c2}^{offset}$ increases for lower temperatures. To find out the upper critical field of the as-synthesized Pb-Bi alloy, the phase diagram is extrapolated to 0K using the Gingberg-Landau model [36]:

$$H_{c2}(T) = H_{c2}(0)\left[\frac{1-t^2}{1+t^2}\right] \tag{6}$$

Where $t=T/T_c$ is the reduced temperature. The fitted data is shown in Fig. 5 by solid black lines. Using the extrapolation of the above-applied model, the upper critical fields are found to be $H_{c2}^{onset}(0) = 3.91$ Tesla and $H_{c2}^{offset}(0) = 3.01$ Tesla. To further calculate the superconducting parameters, the Ginzburg-Landau(GL) formula for coherence length is used, where the upper critical field given by $H_{c2}(0) = \frac{\phi_0}{2\pi\xi_{GL}(0)^2}$, $\phi_0 = h/2e$ is the quantum of magnetic flux and $\xi_{GL}(0)$ is the G-L coherence length at T= 0K. As a result, the G-L coherence length $\xi_{GL}(0)$ is found to be 9.17nm and 10.45nm, for $H_{c2}^{onset}$ and $H_{c2}^{offset}$, respectively. The upper critical field is governed by the Pauli paramagnetic effect ($H_p$ in Tesla) which is limited by the Chandrasekhar-Clogston (or Pauli) paramagnetic limit, $H_p \equiv 1.86T_c$ at T=0K [37]. For the synthesized Pb-Bi alloy, this limit is 16.74Tesla which is almost four times higher than the calculated upper critical field $H_c$. This indicates conventional BCS-type superconductivity present in the sample. However, the upper critical field here reported is higher in comparison to the previous report. Previously, Pb-Bi alloy is reported to have a maximum $T_c \sim 8.7$K with $H_{c2}(0)\sim2$Tesla [15]. Further, we estimate the Ginzberg- Landau penetration depth $\lambda(0)$ using the relation [36]:

$$H_{c1}(0) = \frac{\phi_0}{2\pi\lambda^2(0)}\left[ln\left(\frac{\lambda(0)}{\xi(0)}\right) + 1.2\right] \quad (7)$$

By using $H_{c1}(0) = 291.6$Oe and $\xi_{GL}(0) = 9.17$nm the $\lambda(0)$ is obtained as 8.18nm. The Ginzberg-Landau parameter (κ) is given by $\kappa = \frac{\lambda(0)}{\xi(0)}$. Using the value obtained for $\lambda(0)$ and $\xi_{GL}(0)$, we obtain $\kappa = 0.892 > 1/\sqrt{2}$. It confirms that the synthesized Pb-Bi alloy is a type-II superconductor. The zero temperature thermodynamic critical field is calculated using [36]

$$H_{c2}(0) = \sqrt{2}\kappa H_c(0) \quad (8)$$

and it is found to be 3.1Tesla which is close to the $H_{c2}^{offset}(0)$ value obtained from magneto-transport measurements.

Fig. 6(a) shows the low temperature specific heat capacity data measured in zero field where C/T is plotted against $T^2$. A hump at around $T_c = 8.6$K apparently confirms the bulk superconductivity present in the Pb-Bi alloy. The obtained data is fitted with Sommerfield-Debye expression C/T=$\gamma_n+\beta T^2+\delta T^4$, shown by black solid line in Fig.6 (a). The first term is electronic contribution with Sommerfeld constant $\gamma_n$, following two terms are lattice specific heat taking account two lowest order (up to $T^5$) contributions. The coefficients of Sommerfield-Debye expression have to be determined under the condition that the entropy balance between normal and superconducting states is preserved. The values of Sommerfield coefficient from the fitting of experimental data is obtained as $\gamma_n = 33.19$ mJ/mol-K$^2$ and $\beta = 14.12$ mJ/mol-K$^4$. The Debye temperature ($\Theta_D$) is calculated by using the expression $\Theta_D = (12\pi^4 Rn/(5\beta))^{1/3}$. Here n is the number of atoms per formula unit and R is the gas constant. Using the obtained Sommerfield coefficient $\beta$ the calculated $\Theta_D$ is 111.24K, which is comparable to the value of $\Theta_D$ obtained from B-G fitting of resistivity data. Using the value of $\Theta_D$ and $T_c$, the elctron-phonon coupling constant $\lambda_{el-ph}$ is determined using the McMillan formula [38]

$$\lambda_{el-ph} = \frac{1.04 + \mu^* \ln(\Theta_D/1.45T_c)}{(1 - 0.62\mu^*) \ln(\Theta_D/1.45T_c) - 1.04} \tag{9}$$

Here, the value of screened coulomb potential $\mu^*$ is taken as 0.13 [39,40], which can be assigned any value between 0 to 0.2 for superconductors with $T_c$ below 20K. The calculated value of $\lambda_{el-ph}$ is obtained as 1.36, which indicate Pb-Bi alloy is strongly coupled superconductors, which is in agreement with the previous report [15]. To determine, the presence of bulk superconductivity in the sample, we fit the heat capacity data at the temperatures well below $T_c$ with equation $C = \gamma_{res}T + \beta T^3$ [41] (shown in inset of fig. 6(a)), where $\gamma_{res}T$ and $\beta T^3$ represent the residual electronic specific heat and phonon specific heat respectively. The experimental data is well fitted with the equation, showing that there exist some residual electrons those are not forming the Cooper pairs in zero temperature limit. The obtained values of $\gamma_{res}$ and $\beta$ are found to be 2.52 mJ/mol-K$^2$ and 20.19 mJ/mol-K$^4$. The value of $\gamma_{res}$ is 7.6% of $\gamma_n$ which is reasonably small. This shows the observed superconductivity in the synthesized Pb-Bi alloy is bulk superconductivity.

Fig. 6(b) show the specific Heat capacity measurements performed at fields 0T, 0.1T, 0.2T, 0.3T, 0.4T and 0.5T. As the field increases, the observed hump height decreases along with the decreasing $T_c$. The variation of $T_c$ with applied field is shown in the inset of Fig.6 (b). The obtained data is fitted with equation 6, resulting in upper critical field $H_{c2}(0) = 1.85$T, which is comparable to the magnetization results. For further investigation, Muon spin rotation (µSR) experiments are warranted.

The presence of strong coupling can be determined by the theoretical fitting of the temperature dependence of electronic heat capacity in the superconducting state as shown in Fig. 6(c). The electronic heat capacity is obtained by subtracting the phononic contribution from measured heat capacity [41] and the same is plotted against the temperature in fig. 6(c). Entropy balance is maintained while subtracting the phononic contribution as revealed by nearly similar area below and above the zero base line in the superconducting state, which shows that the entropy is balanced, consistent with the thermodynamics [42]. Interestingly, the electronic heat capacity can not be fitted for the single superconducting gap as shown by the blue curve in Fig. 6(c). A clear anomaly is observed at around 6.2K, which is similar to as observed for some two gap superconductors [43,44]. Thus, the heat capacity in the superconducting state is tried to fit with two-gap equation, and is found to be well fitted as shown by green curve in fig. 6(c). This shows the possibly two superconducting gaps are present in the synthesized Pb-Bi alloy. The obtained superconducting gap values are $\Delta_1$ = 2.05 meV and $\Delta_2$ = 0.83 meV. These values close to that the observed values in ref. 15, involving only the single superconducting gap. The corresponding values of the parameter $\alpha$ are obtained for each superconducting gap using the formula $\alpha = \frac{\Delta_0}{k_B T_c}$. The obtained values are $\alpha_1$ = 2.77 and $\alpha_2$ = 1.12 corresponding to superconducting gaps $\Delta_1$ and $\Delta_2$. The first parameter $\alpha_1$ is higher than the values reported in ref. 15, which is reasonable as the observed $T_c$ is also higher than the reported ones. This value of parameter $\alpha$ is well above the maximum value for BCS weakly coupled superconductors [39,40,41]. This also confirms strongly coupled superconductivity in the synthesized Pb-Bi alloy sample. For the small gap, the characteristic ratio $\alpha_2$ is within the BCS weak coupling limit 1.77, which is typical for a weak SC condensate in a multigap superconductor [45]. Above analysis shows Pb-Bi alloy have possibly two-gap superconductivity, however, more studies on pure and single crystals of Pb-Bi alloy are warranted.

The Bi in single crystal form shows superconductivity at extremely low temperatures (~mK) [46], whereas the Bi nanoparticle exhibits $T_c$ up to 8.75K [47]. The GL coherence length for our sample is smaller in comparison to superconducting Bi nanoparticles but comparable to that of Pb-Bi alloy reported by Gandhi et al [15]. For, pure Pb the $T_c$ is 7.2K [48] which is below the observed $T_c$ ~ 9K of synthesised Pb-Bi alloy. Also, PXRD analysis does not show any Pb impurity peak. In specific heat capacity measurement, the observed anamoly can not be due to Pb as the observed superconducting gap for elemental Pb ($\Delta(0)$ = 1.43meV) [49] lies in between the two gaps observed for Pb-Bi alloy. Thus, in the synthesized Alloy of Pb-Bi, the observed superconductivity is due to Pb$_7$Bi$_3$ phase. We have achieved a higher T$_c$ and upper critical field for the Pb-Bi alloy compared to the previous reports.

**Conclusion**

To summarise, here we report the successful growth of crystalline Pb-Bi alloy following a simple single-step solid-state reaction route, which contains two phases i.e., Pb$_7$Bi$_3$ and Bi and no Pb impurity is observed as characterized by XRD and EDAX measurements. The electronic transport ρ-T plot shows that the synthesized Pb-Bi alloy has T$_c$ at 9K. The magnetotransport measurement was performed and the superconductor phase diagram extrapolation resulted in the upper critical field $H_{c2}^{onset}$ to 3.9Tesla at 0K. The magnetization measurement shows PME due to the presence superconductor-normal conductor interface. The wide opening in the M-H plot shows type-II superconductivity which is further substantiated by calculating critical superconducting parameters viz. coherence length, penetration depth, G-L κ paramter etc. By comparing the superconducting parameters obtained by the phase diagram, we conclude that the superconductivity shown by the synthesized alloy is due to the majority Pb$_7$Bi$_3$ phase. The bulk superconductivity in the synthesized Pb-Bi Alloy is endorsed by the heat capacity measurement, which shows it to be a strongly coupled superconductor with a possible two-gap superconductivity.

**Acknowledgment**

The authors would like to thank the Director of the National Physical Laboratory (NPL), India, for his keen interest in the present work. Also, the authors would like to thank CSIR and UGC, India, for the research fellowship, and AcSIR-NPL for Ph.D. registration. The authors would like to acknowledge Dr. J. S. Tawale for SEM/EDAX and Dr. Pallavi Kushwaha for MPMS measurements. This work is supported by NPL in-house project numbers OLP 240832 and OLP 240232. Heat Capacity measurements are supported as part of the μ-ATOMS, an Energy Frontier Research Center funded by the U.S. Department of Energy, Office of Science, Basic Energy Sciences under award DE-SC0023412.

**Table 1**
**Parameters obtained from Rietveld refinement**

| $Pb_7Bi_3$ Cell Parameters | Bi Cell Parameters | Refinement Parameters |
|---|---|---|
| Cell type: Hexagonal<br>Space Group: P63/mmc<br>Lattice parameters:<br>a=b=3.5051((1)Å<br>& c=5.7984(2) Å<br>$\alpha=\beta=90°$ & $\gamma=120°$<br>Atomic co-ordinates:<br>Pb (0.0770(7), 0.1541(3),0.25)<br>Bi (0.5522(3), 0.2222(3),0.75)<br>Frac(%) 45.51 | Cell type: Rhombohedral<br>Space Group: R-3m (166)<br>Lattice parameters:<br>a=b=4.5435(9)Å<br>& c=11.8694(5) Å<br>$\alpha=\beta=90°$ & $\gamma=120°$<br>Atomic co-ordinates:<br>Bi(0,0,0.2321)<br><br>Frac(%) 54.49 | $\chi^2$=3.7<br>$R_p$=16.0<br>$R_{wp}$=20.2<br>$R_{exp}$=9.88 |

**Figure Captions**

**Figure 1:** Heat treatment of the Pb-Bi alloy and inset shows the as-grown polycrystalline sample.

**Figure 2: (a)** Rietveld refinement of PXRD pattern of synthesized Pb-Bi alloy. **(b)** The rhombohedral unit cell of Bi (in left) and hexagonal unitcell of $Pb_7Bi_3$ (in right) drawn using VESTA software. **(c)** Field Emission Scanning Electron Microscope (FESEM) image of grown Pb-Bi alloy and the inset shows the EDAX spectra of synthesized Pb-Bi alloy with elemental composition.

**Figure 3: (a)** Temperature-dependent resistance of Pb-Bi alloy at zero field and the inset shows the enlarged image near the transition temperature. **(b)** Show the temperature-dependent resistivity of Pb-Bi alloy for various applied fields. **(c)** The magnetic field dependence of the resistivity $\rho(H)$ in Pb-Bi alloy in temperature range 2-10K.

**Figure 4: (a)** Illustrates the temperature-dependent magnetization of Pb-Bi alloy measured in ZFC and FC protocol at 10 Oe. **(b)** Isothermal magnetization at 2K and 5K in five quadrants.

**Figure 5:** Variation of the critical field ($H_{c2}$) with reduced temperature for Pb-Bi alloy and fit to equation 6. The inset depicts the $H_{c1}$ variation with reduced temperature fitted within G-L formalism by equation 5.

**Figure 6: (a)** Specific Heat capacity C/T vs. T plot for Pb-Bi alloy at 0Tesla. The inset shows C/T vs. $T^2$ linearly fitted with C/T=$\gamma_{res}+\beta T^2$. **(b)** The specific heat data C/T vs. T plot at different applied magnetic fields and the inset shows variation of $T_c$ with applied magnetic field. **(c)** Electronic heat capacity for superconducting state fitted with single gap and two gap superconductor equation.

Figure 1:

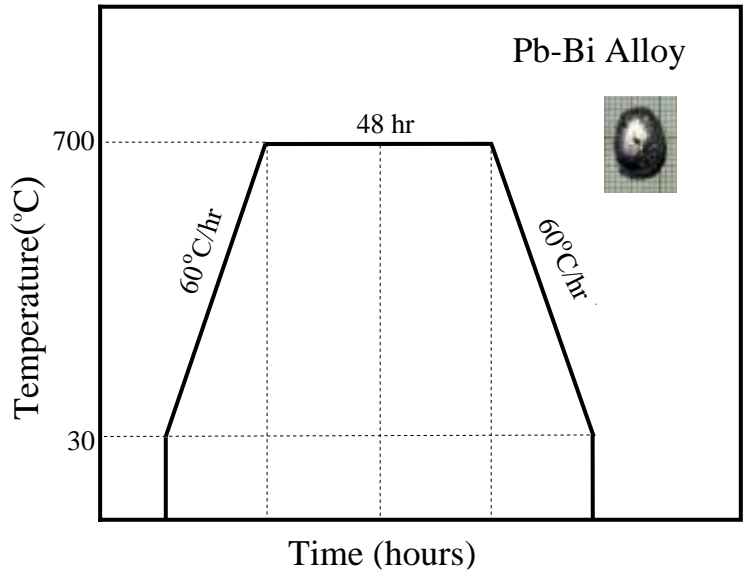

Figure 2(a):

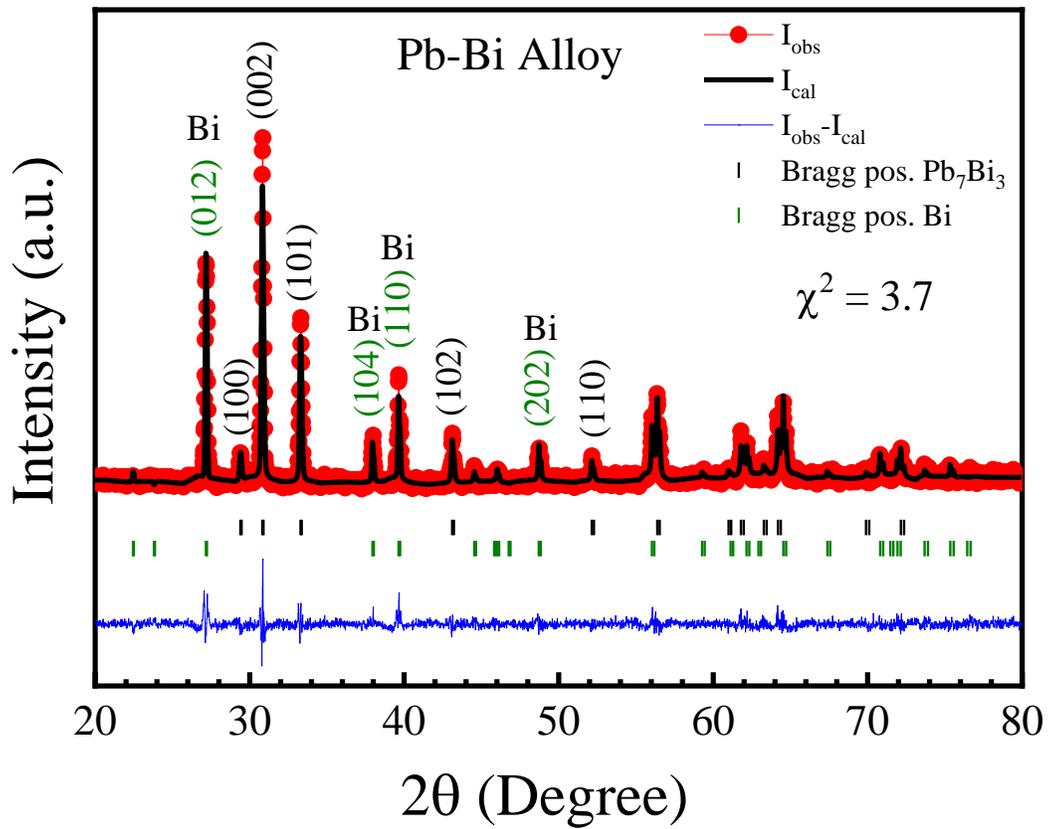

Figure 2(b):

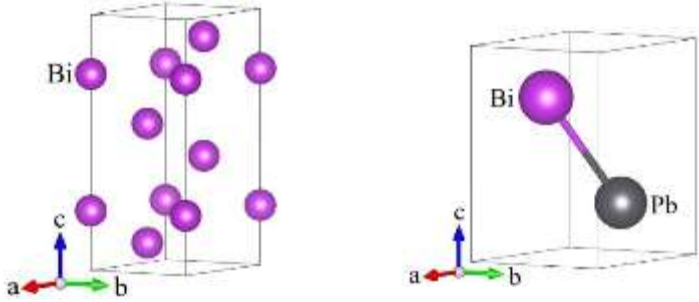

Figure 2(c):

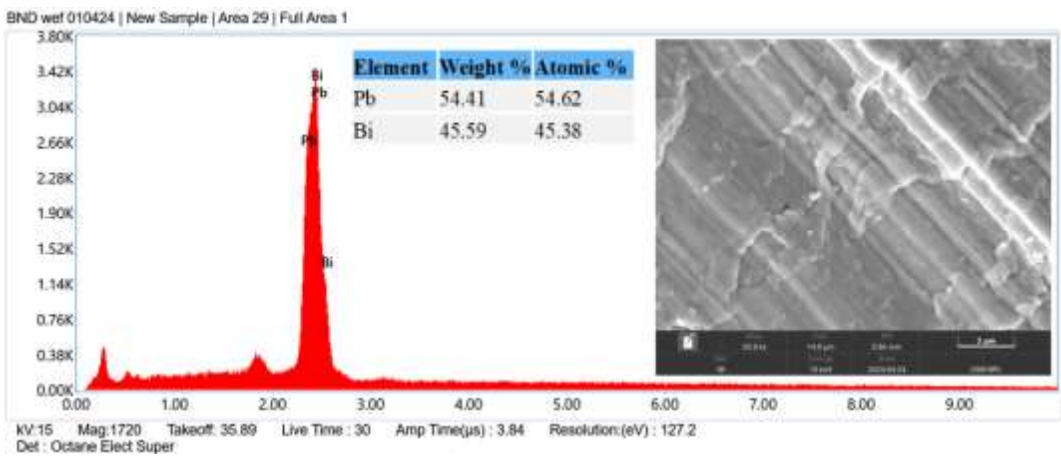

Figure 3(a):

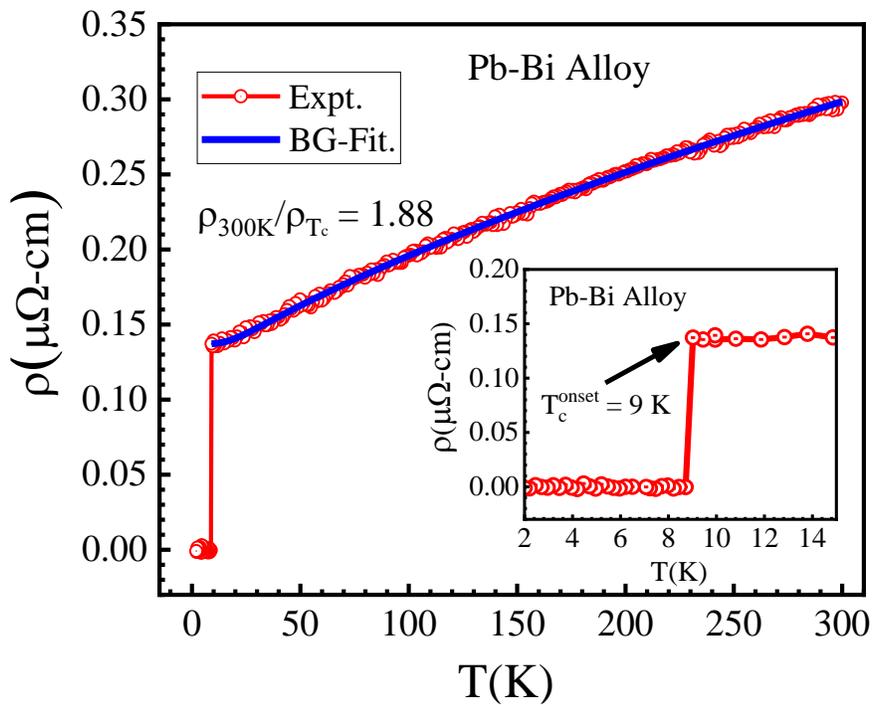

Figure 3(b)

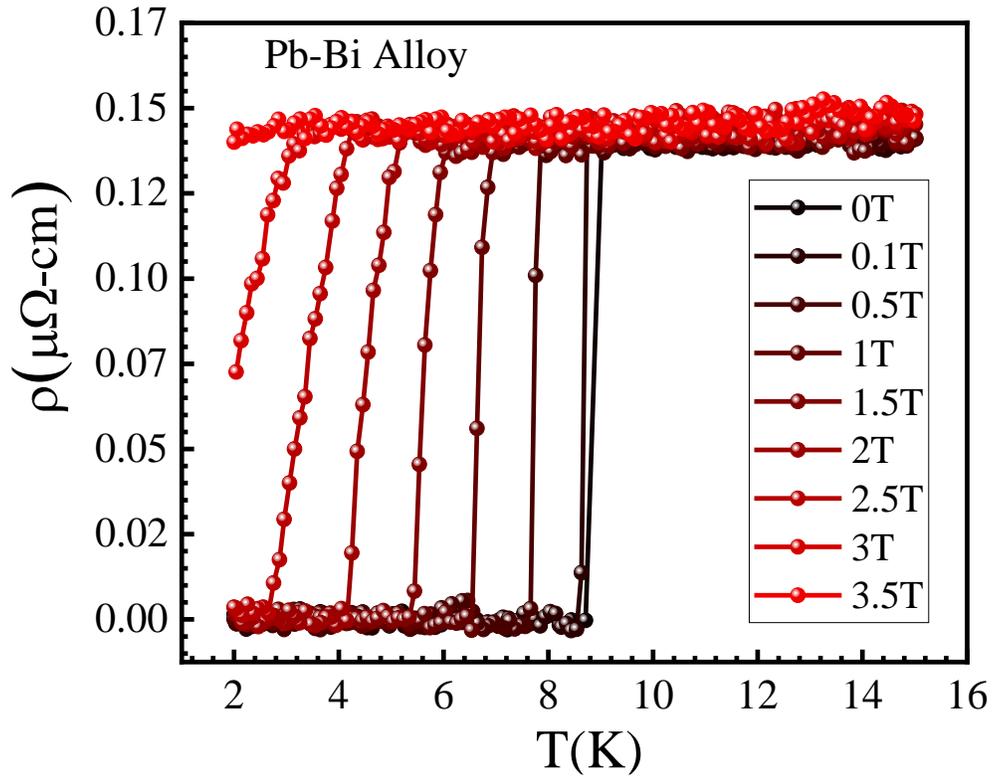

Figure 3(c)

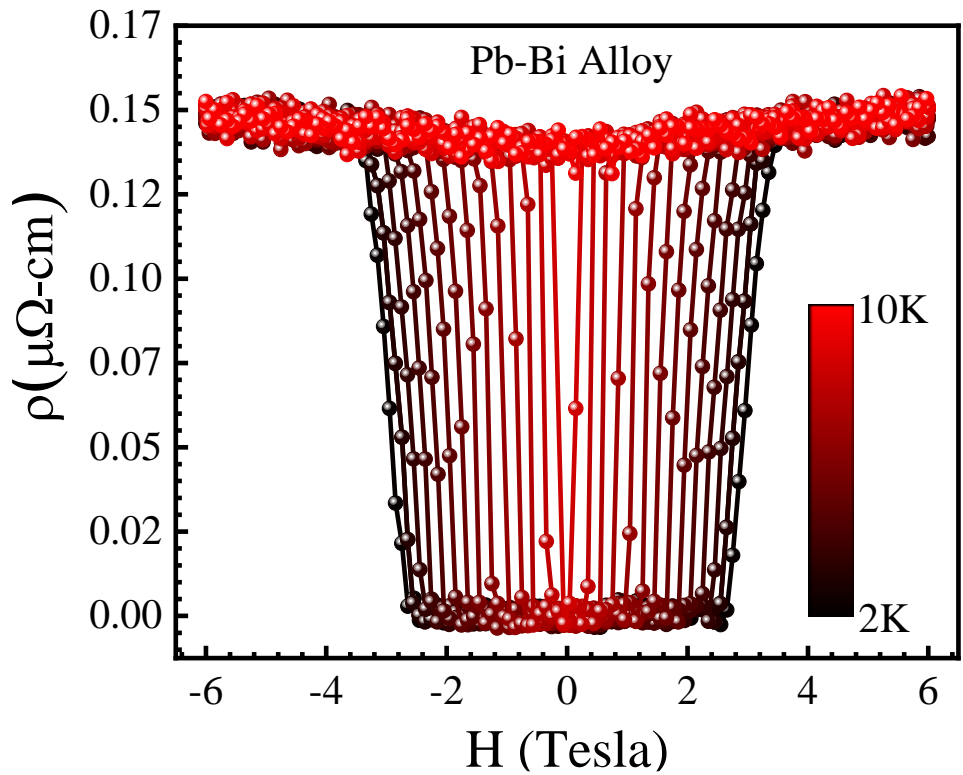

Figure 4(a)

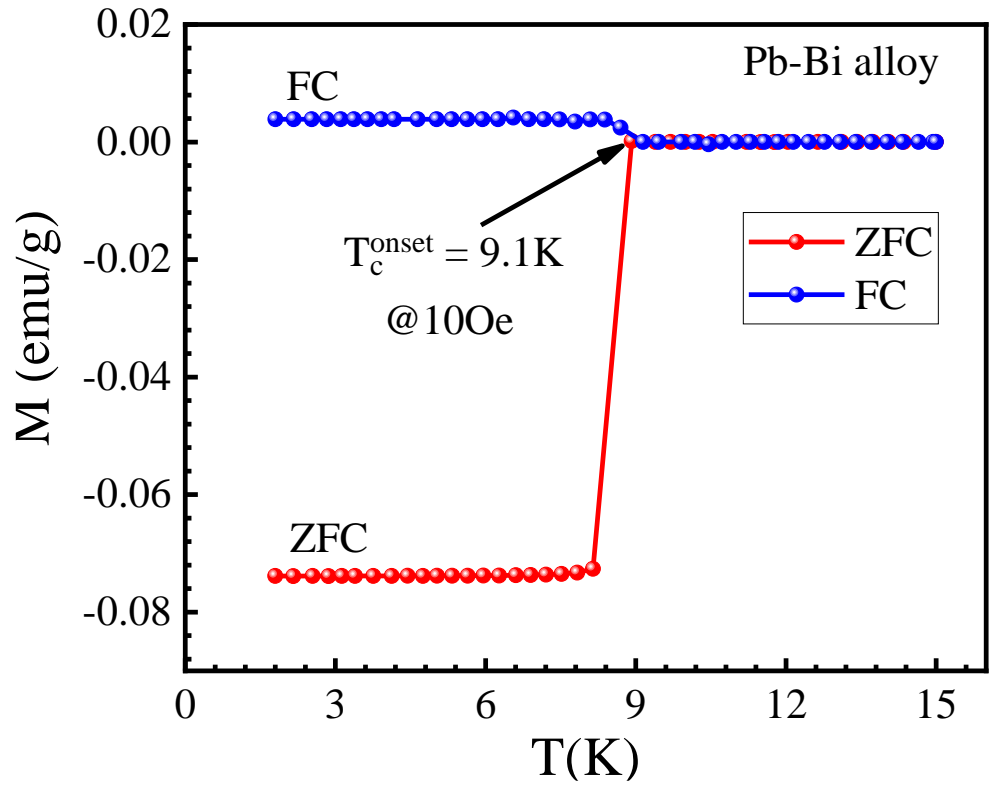

Figure 4(b)

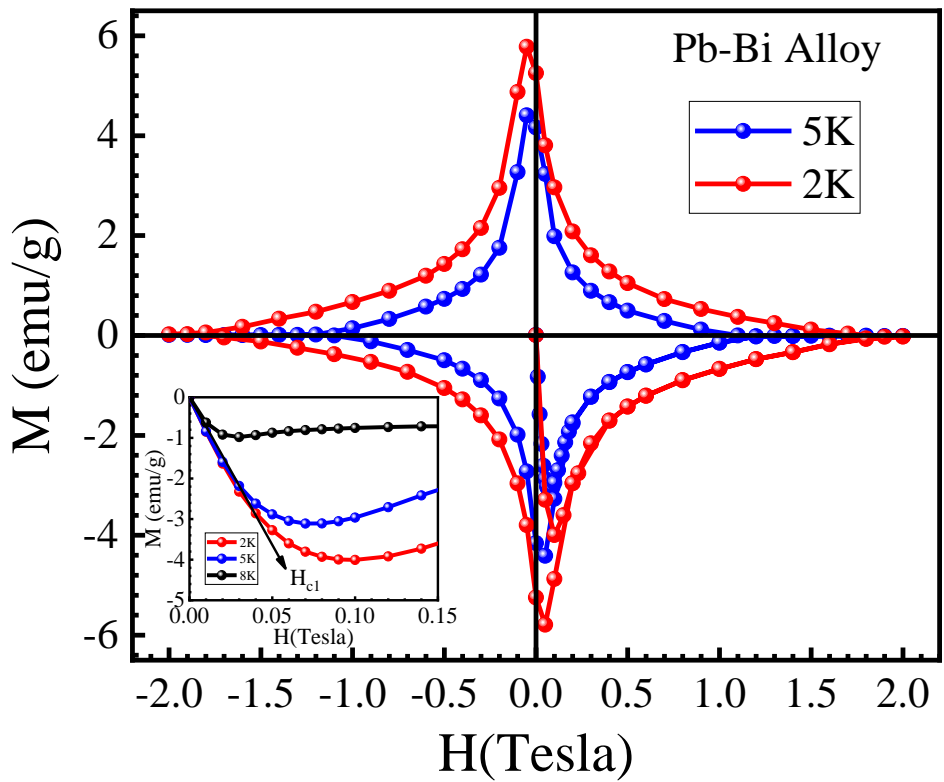

Figure 5

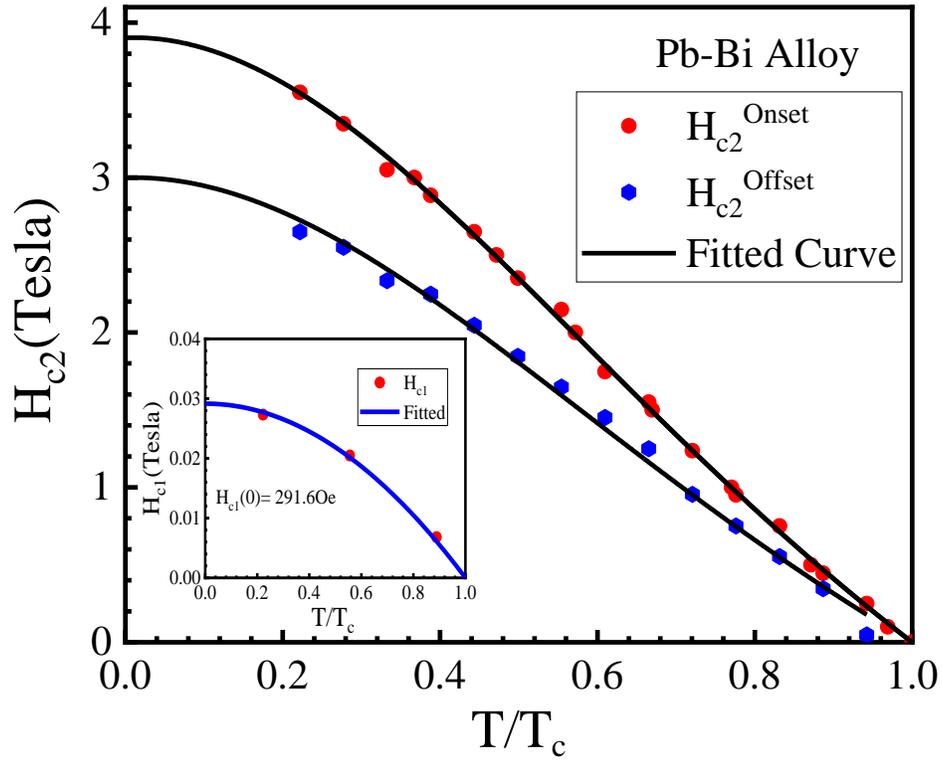

Figure 6(a)

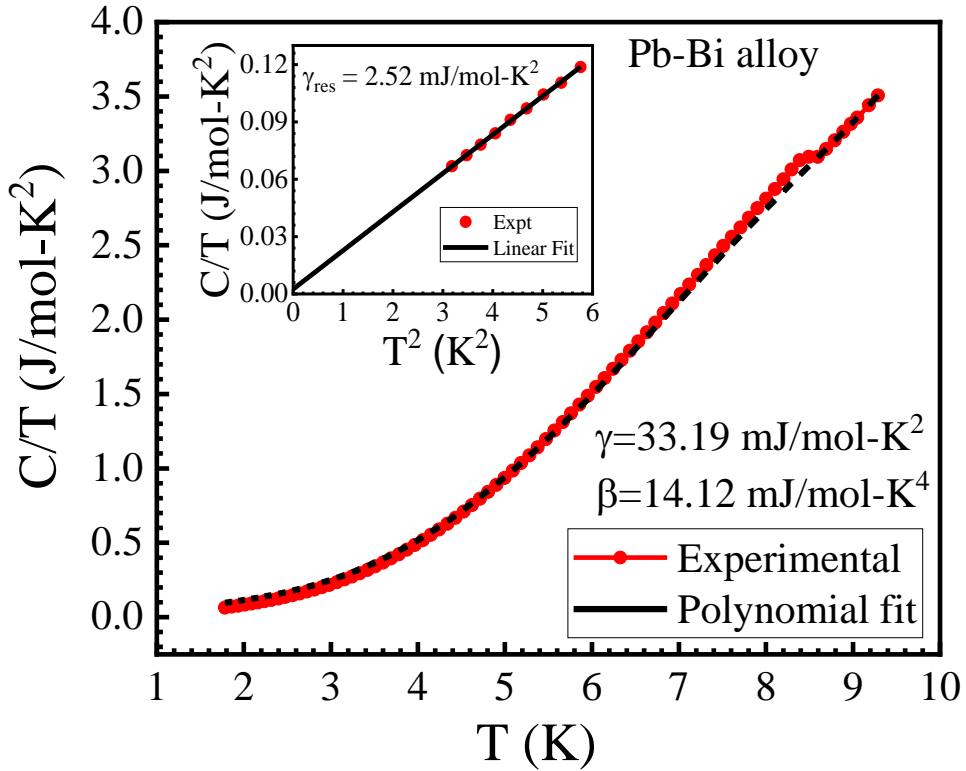

Figure 6(b)

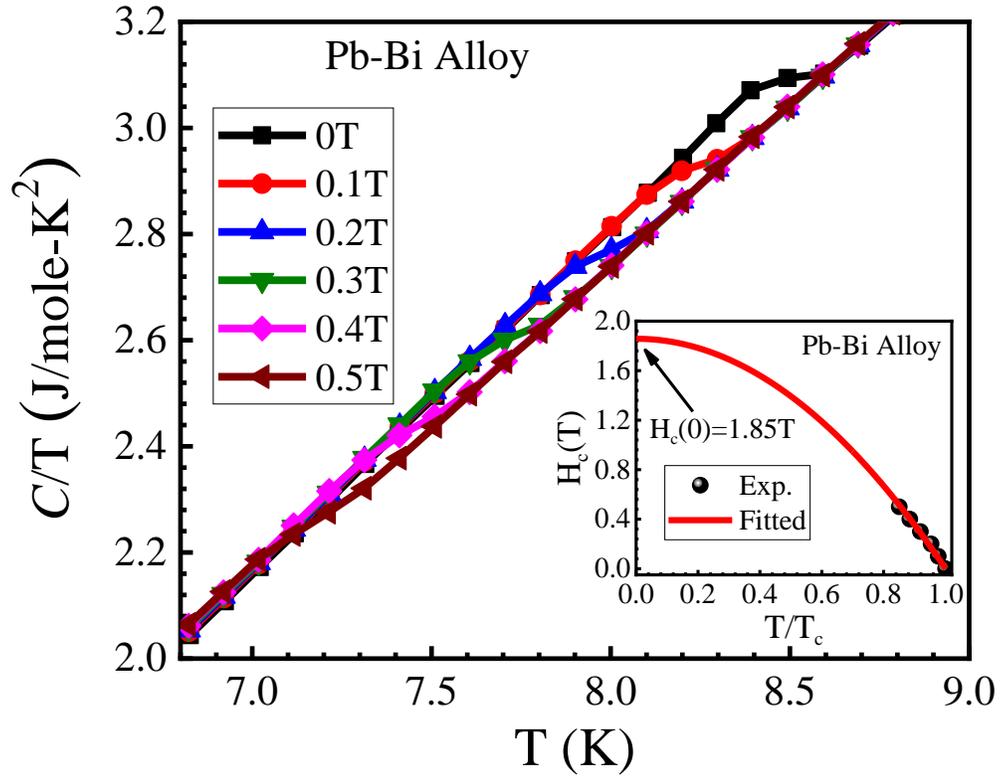

Figure 6(c)

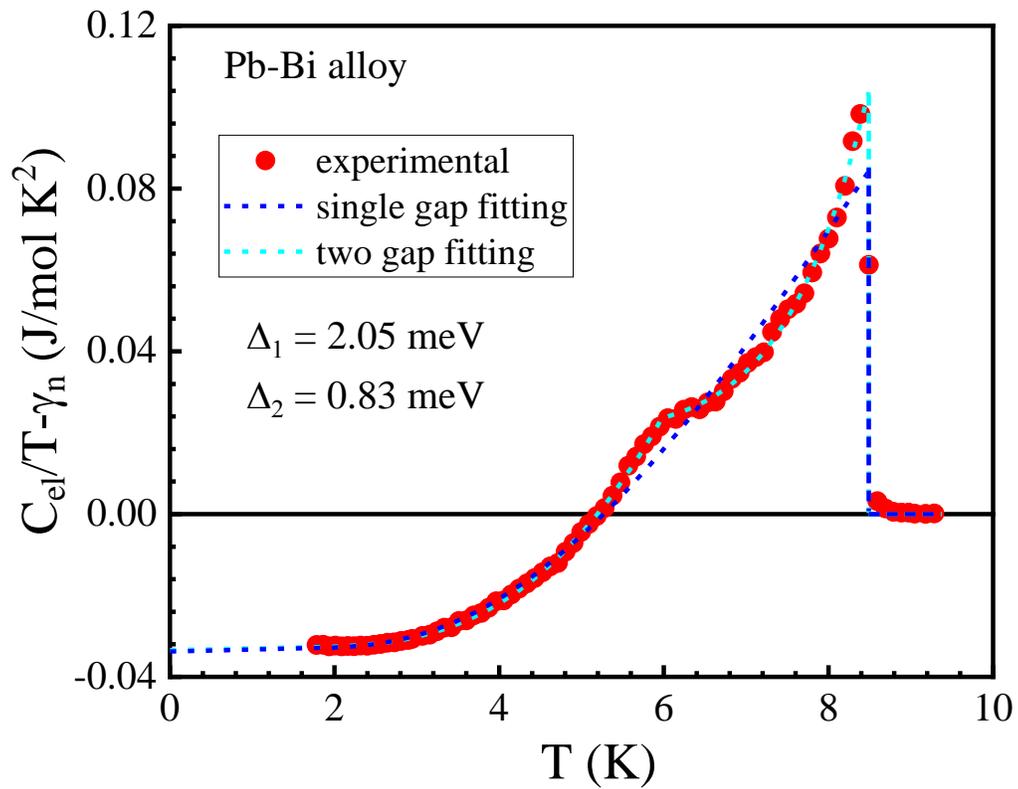